\documentstyle[12pt]{article}
\input psfig

\def\reference{\parskip 0pt\par\noindent\hangindent 0.5 truecm}
\def\kms{km ${\rm s}^{-1}$}
\begin{document}
%
%
\title{The impact of the early stages of radio source evolution on the ISM of
the host galaxies}
%


\author{R. Morganti,$^{1}$ 
 C.N., Tadhunter,$^{2}$ 
 T.A. Oosterloo,$^{1}$
 J. Holt,$^{2}$ \and
 A. Tzioumis,$^{3}$ 
 K. Wills,$^{2}$
} 

\date{}
\maketitle

{\center
$^1$ ASTRON, PO Box 2, 7990 AA Dwingeloo, The Netherlands\\
morganti@astron.nl\\[3mm] 
$^2$ University of Sheffield, Hicks Building, Hounsfield Road, Sheffield, S3 7RH,
UK\\
c.tadhunter, j.holt,
k.wills@sheffield.ac.uk\\[3mm]
$^3$ ATNF, PO Box 76, Epping NSW 1710, Australia\\
atzioumi@atnf.csiro.au\\[3mm]
}

%
\begin{abstract}

The study of both neutral and ionized gas in young radio sources is
providing key information on the effect the radio plasma has on the
ISM of these objects.  We present results obtained for the compact
radio sources PKS~1549--79, 4C~12.50 and PKS~1814--63 and for the
intermediate-size radio galaxy 3C~459.  At least in the first two,
low ionisation optical emission lines and HI absorption appear to be
associated with the extended, but relatively quiescent, dusty cocoon
surrounding the nucleus.  The [OIII] lines are, on the other hand,
mostly associated with the region of interaction between the radio
plasma and the ISM, indicating a fast outflow from the canter.  A case
of fast outflow (up to $\sim 1000$ \kms) is also observed in HI in the
radio source 4C~12.50. As the radio source evolves, any obscuring
material along the radio axis is swept aside until, eventually,
cavities (of the same kind as observed e.g.\ in Cygnus ~A) are
hollowed out on either side of the nucleus. We may witness this phase
in the evolution of a radio source in the radio galaxy 3C~459.

\end{abstract}

{\bf Keywords:}

\bigskip

%
%

\section{Introduction}

Although in the recent years important progress has been made in the
understanding of the physics of AGN, a number of main questions
still remain open. One of them is the early evolution of powerful
radio galaxies.  To know more about this phase is important not only
from the point of view of the detailed phenomenology of these objects,
but also for understanding the evolution of massive galaxies.  In
general, galaxies in their early stage of radio activity are likely to
have their nuclear regions enshrouded in a cocoon of material left
over from the accretion event(s) that trigger the activity. As the
activity evolves, and the radio jets expand, the ISM and dust along
the radio axis will be swept from the nuclear regions by jet-cloud
interactions and/or quasar-induced winds. Thus, outflows of gas can be
significant in this phase.  These processes have been suggested to
profoundly affect even the star formation history in luminous galaxies
(Silk \& Rees 1998). Thus, the study of the early-phase of radio
activity and its effect on the galaxy medium has broad implications.

As described in many contributions in these Proceedings, both neutral
and ionized gas are clearly present around young radio sources.  The
study of this gas has a key role in the investigation of the
evolutionary scenario described above. Spectral line observations
can, in fact, provide information on the gas kinematics where
signatures of interaction with the radio plasma can be found.  Here we
present the results (some still preliminary) obtained in the study of
the HI in three compact radio sources (PKS~1549--79, 4C12.50 and
PKS~1814--63) and one intermediate-size radio galaxy (3C~459).  While
the HI detected in compact radio sources is very often interpreted as
due to a circum-nuclear disk/torus, it is now clear that more
complicated situations are present (see e.g. Morganti 2002 and
ref. therein) and that neutral gas can also trace the presence of
(sometimes extreme) interactions.  The objects described here are also
part of a detailed study of their ionized gas. This information is
crucial for the correct interpretation of the kinematics of the
neutral hydrogen, as it will be illustrated below.

\begin{figure}
\centerline{\psfig{file=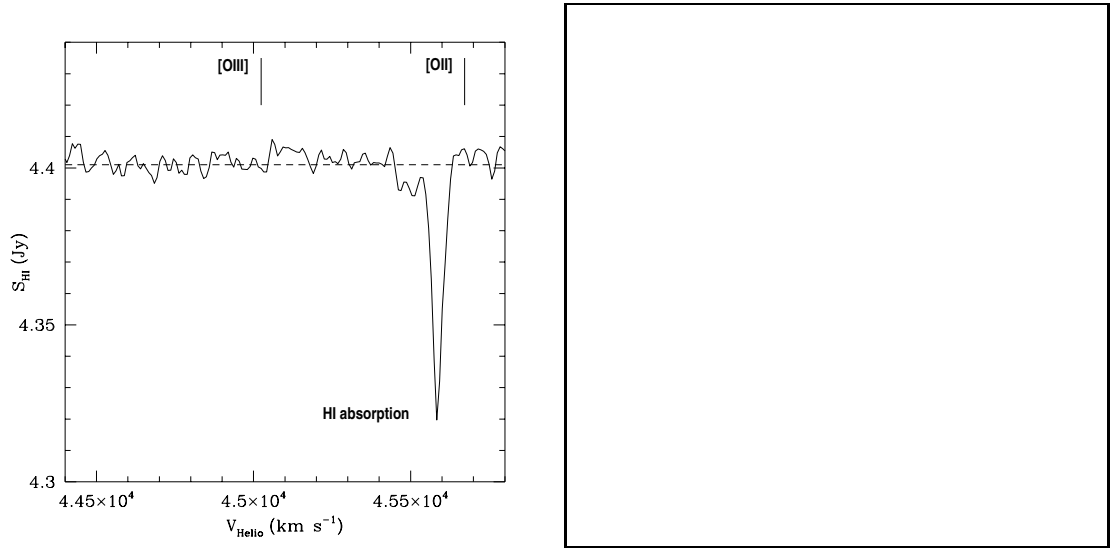,height=7cm}
\psdraft
\psfig{file=Morganti.fig1.ps,height=7cm}
\psfull}
\caption{{\sl Left:} profile of the HI absorption in PKS~1549--79 
(from Morganti et al. 2001). The
velocities derived from the optical emission lines are indicated. {\sl
Right:} continuum radio emission on the VLBI scale from King 1996.}
\end{figure}

\section{The case of PKS~1549--79}

In the above context, we have recently studied the southern radio galaxy
PKS~1549--79.  This is a compact radio source with a one-sided jet on
the VLBI scale (see Fig.~1).  The overall size is about 350 pc 
\footnote{We use here $H_\circ = 50$ km/s Mpc$^{-1}$ and
$q_\circ = 0$}.  The radio structure suggests that the radio jet axis
is aligned close to our line of sight (King 1996).  Because of these
characteristics, it was quite surprising to detect HI absorption in
this object (Morganti et al.  2001).  The absorption has a peak
optical depth of $\sim 2$\% (see Fig.~1) and a column density of $\sim
4\times 10^{20} T_{\rm spin}/100K$ cm$^{-2}$.

The optical spectrum of PKS~1549--79 is rich in strong emission lines
(Tadhunter et al. 2001).  Interestingly, the high ionisation ([OIII],
[NeIII] and [NeV]) lines are unusually broad (FWHM $\sim 1200$ \kms)
and blue-shifted by 600 \kms\ relative to the much narrower low
ionisation (i.e.  [OII] and [OI]) lines.  The remarkable result is
that the velocity of the HI absorption agrees quite well with that
derived from the low ionisation optical lines (see Fig.~1).

Our interpretation of all this (Tadhunter et al.  2001) is that {\sl
the low ionisation emission lines and HI absorption are associated
with the extended, but relatively quiescent, dusty cocoon surrounding
the nucleus}, whereas the [OIII] lines are associated with the region
of interaction between the radio plasma and the ISM and indicating a
fast outflow from the canter.  We cannot yet fully exclude the
possibility of starburst or AGN winds as responsible for the
interaction. However, the common presence of broad (forbidden) lines
in young, powerful radio sources (Gelderman \& Wittle 1994) seems to
point to the radio jet as the likely candidate to produce the
interaction.  A cartoon of the proposed model is presented in Fig.~3
in Holt et al. (these Proceedings) and described in more details in
Tadhunter et al. (2001).  The interaction of the radio plasma with
this gas would explain also the larger FWHM of the high ionisation
lines.

\begin{figure} 
\centerline 
{\psfig{file=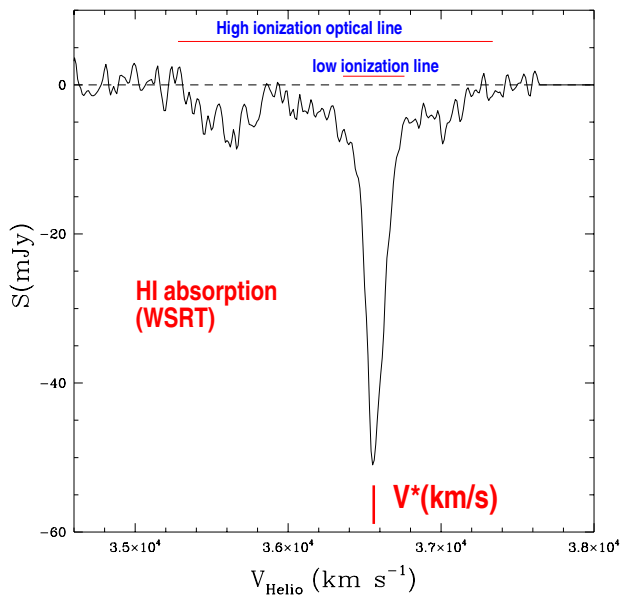,height=8cm}}
\caption{Profile of the HI absorption in 4C~12.50 as obtained from WSRT
 observations. The systemic velocity from stellar absorption lines (Grandi
 1977) is marked. The range in velocity of two of the optical emission line components 
is also indicated (see Holt et al. these Proceedings for details).}

 \end{figure}

\section{The gas outflow in 4C~12.50}

One other object known to have optical characteristics similar to
PKS~1549--79, i.e. two redshift systems and very broad optical lines,
is the GigaHertz Peaked galaxy 4C~12.50 (see Grandi 1977). New optical
spectra obtained for this object are described in detail in Holt et
al. (these Proceedings). Three gaussian components are needed 
to fit the strong [OIII] lines. One of them is a low ionisation
component and it is cantered on the systemic velocity (that is derived
by Grandi (1977) from the stellar absorption lines). The other two
components show much higher ionisation and are highly blue-shifted.
The situation appears, therefore, similar to  PKS~1549--79.

The presence of HI absorption in this galaxy was already known from
Arecibo observations (Mirabel 1989).  However, new WSRT observations
show that this absorption is actually {\sl much broader than previously
known}.  As shown in Fig.~2, an extremely broad absorption system is
detected together with a (relatively) narrow and deeper component.  This
(relatively) narrow absorption ($\tau \sim 0.01$, column density $\sim 2
\times 10^{20} T_{\rm spin}/100K$ cm$^{-2}$) has a velocity similar to
the systemic velocity detected from the stellar component and similar to
the low ionisation component.  Also in HI the situation appears,
therefore, similar to PKS~1549--79.  On the other hand, the overall
system of shallower HI absorption ($\tau \sim 0.002$, column density
$\sim 10^{20} T_{\rm spin}/100K$ cm$^{-2}$) is about 2000 km/s broad
with a very large blue-shifted wing (more than 1000 \kms) as well as a
redshifted wing of few hundred \kms.  This broad absorption may
represent an outflow of gas due to the interaction of the radio plasma
with the dense ISM (e.g.  gas in a cocoon around or entrained by the
radio jet?).  The blue-shifted component would be then associated to the
jet pointing toward us (and the opposite for the redshifted component).
VLBI observations have now been obtained for 4C~12.50 to investigate the
detailed morphology of the HI absorption on the pc scale. 
The range of velocity of the broad HI absorption appears consistent with
one of the high ionisation components found in the emission lines (see
Holt et al.  these Proceedings).  In this galaxy, therefore, the gas
outflow would be seen both in the ionized and in the neutral gas. 
The scenario presented for PKS~1549-79 appears to explain also
what is observed in 4C~12.50 with the radio plasma being in the process
of clearing its way through the rich ISM.  It is however not yet clear
how gas associated to such a fast outflow can remain neutral and not
fully ionized.  A fast outflow seen in neutral hydrogen is, however, not
unique to 4C~12.50.  We have at least an other case in the radio-loud
Seyfert~2 galaxy IC~5063 (Oosterloo et al.  2000).  In this case, the
radio plasma ejected from the nucleus is believed to interact directly
with a molecular cloud.

\begin{figure}
\centerline{
\psfig{file=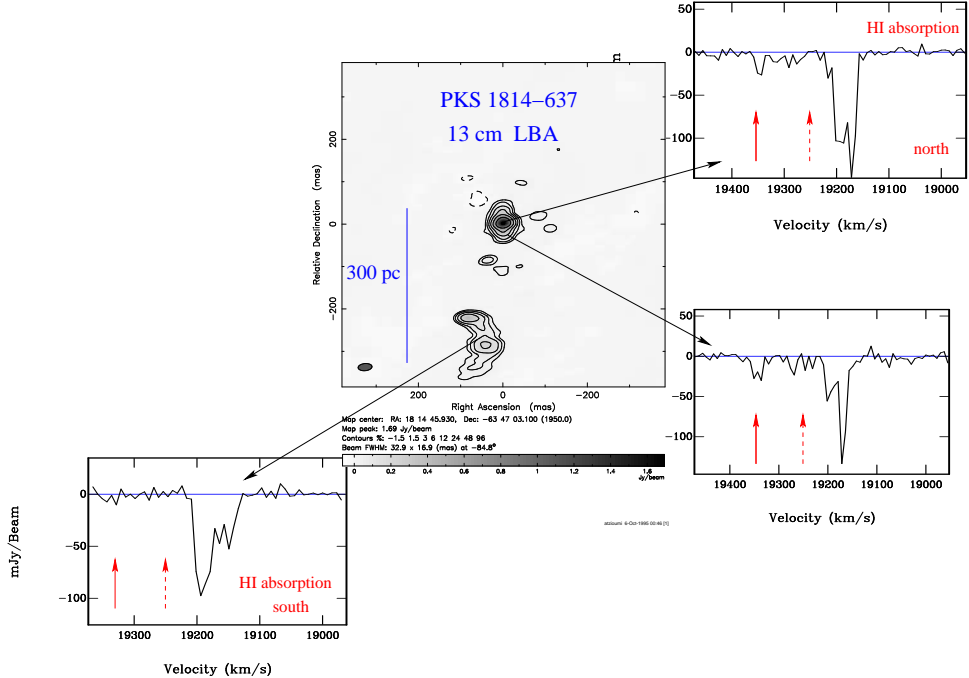,height=9cm}}
\caption{Continuum image of PKS~1814-63 at 13cm and spectra of the HI
absorption obtained at three different locations in the source. The vertical
lines indicate different values quoted for the  systemic velocity of this galaxy.}
\end{figure}

\section{A promising case: PKS~1814--63?}

An other promising case is the compact (steep-spectrum) radio galaxy
PKS~1814--63.  HI absorption was detected from low-resolution
observations (Morganti et al.  2001) and then followed up by LBA
observations.  Thus, unlike the sources described above, for
PKS~1814--63 the distribution of the HI on the VLBI scale is known.  On
this scale, we detect HI absorption against the entire radio source
(about 350 pc, see Fig.3).  Unfortunately, the optical redshift is still
quite uncertain (two of the values found in the literature are indicated
in Fig.3) due to the poor quality of the available optical spectra. 
However, new high-quality optical spectra have been obtained very
recently (but not yet fully analysed) and they show  a
complex kinematics of the ionized gas.  Until, more information from the
optical spectra is obtained, the interpretation of the kinematics of the
neutral gas is still ambiguous. 

The deep absorption (that has a very high value of optical depth $\tau
\sim 0.2$, Morganti et al.\ 2001) could be due to a large-scale gas system. 
The fact that this absorption seems to have more components and their
relative ratio changes for different positions against the radio source,
could be connected with the presence of sub-structures in the gas
screen.  However, if the redshift of 19350 \kms\ is confirmed (drawn line
in Fig.~3, see Morganti et al.\ 2001 for a discussion), this gas could
also be undergoing a mild outflow perhaps connected to the expansion of
the source in the surrounding medium.  If this is the case, we expect to
see kinematical signatures also in the optical spectrum (like in
PKS~1549--79 or 4C~12.50).  Finding the actual systemic velocity will also
help us in interpreting the shallow, broader component.  This is
detected only against the norther part of the radio source and could be
due to a circumnuclear-disk structure, although it is not clear yet from
the available radio data whether this component actually correspond to
the core. 

Finally, it is clear that in the study of PKS~1814--63 we are strongly
limited in the interpretation of the HI absorption by the lack of more
detailed optical information.

\begin{figure} \centerline {\psfig{file=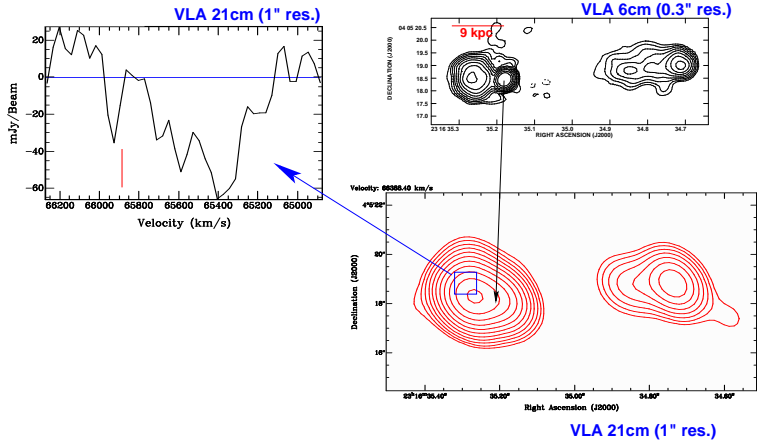,width=10cm}}
\caption{Radio continuum image of 3C~459 from VLA 6cm ({\sl Top panel})
from Morganti et al.  (1999) and from the line-free channels of the VLA
21cm ({\sl Lower panel}).  {\sl Left panel:} HI absorption spectra.  The
systemic velocity (from Spinrad et al.  1985) is indicated.}
\end{figure}

\section{The intermediate scale: 3C~459 }

The above examples indicate that, at least in some cases, the obscuring
material may cover large fraction of the sky as seen by the central
source.  As the radio source evolves, any obscuring material along the
radio axis is likely to be swept aside and dissipated by jet-cloud
interactions or quasar-induced winds until, eventually, cavities are
hollowed out on either side of the nucleus.  Such cavities have been
detected, for example, in the powerful extended radio source Cygnus~A
(Tadhunter et al.  1999).  Before this stage is reached a substantial
amount of material may be present along the radio axis, but at larger
distances from the nucleus. 

An intermediate case that we are studying in some detail is 3C~459.  This is a
small and very asymmetric radio source.  Its total size ($\sim 40$ kpc) is
much larger than the typical CSS.  HI absorption has been detected in this
source (Morganti et al.  2001) and new higher resolution observations (A-array
VLA) tentatively show that most of the absorption could be located against the
eastern radio lobe (see Fig.~4) at about 9 kpc from the radio core.  Thus,
this could be a case of intermediate-size object where we can study the
evolution of the gas after the most compact phase of the CSS is past but when
the source still has a galactic size.  Furthermore, there could also be a
relationship between the small size of the eastern component and the presence
of a large amount of gas on that side of the source. 

It is important to point out that the two compact objects mentioned
above (PKS~1549--79 and 4C~12.50) as well as 3C~459 are far-IR bright
galaxies and show, in their optical spectra, the presence of a young
stellar population (we have no yet information about PKS~1814-63).
This is a particularly interesting class of objects where HI
absorption is often detected. This could therefore point to the
possibility that these objects are actually ``special'' because of
their particularly rich ISM.

\section{Summary}

We have shown as (at least some) compact radio galaxies can have
different gas components with very different kinematics.  The neutral
hydrogen does not appear to be associate only to circum-nuclear
disk/tori (as often claimed), but instead there is evidence of the
presence of more diffuse and quiescent halos in which these young sources
are embedded.  The radio jets are expected to clear their way through
the rich ISM.  As indication of this, high speed outflow are detected
not only in the ionized gas but, quite surprisingly, also in neutral
hydrogen in the case of 4C~12.50. 

The kinematics of the neutral hydrogen alone can be sometimes difficult
to be unambiguously interpreted. However,  it can provide crucial information
on the condition, morphology and kinematics of the gas in the central
regions of radio sources once it is combined with the results from the
ionized gas.  This has been clearly shown for the case of the third
radio galaxy presented, PKS 1814-63.  Despite the complex and intriguing
HI absorption detected, the interpretation is still uncertain due to the
lack of high quality optical spectra.  Finally, we have presented a case
of intermediate-size radio source (3C~459) where we may witness the
next step of the evolution of the source. 

In how many young radio sources do we expect to see the effect of the
radio plasma making its way through the ISM and producing gas
outflows?  We do not have yet enough information available, especially
in the optical (e.g.\ accurate measurement of the redshift), to
estimate this in a reliable way.  The group of radio galaxies that are
far-IR bright and that show a young stellar population in their
optical spectra seems, however, to be the most likely to show these
extreme effects and outflows.





\section*{References}





\reference Axon D.J., Capetti A., Fanti R., Morganti R. et al. 2000, AJ 120, 2284;
\reference Gelderman R., Whittle M. 1994, ApJS 91,491;
\reference Grandi S.A. 1977, ApJ, 215, 446;
\reference King E. 1996 PhD Thesis Univ. of Tasmania;
\reference Maloney P.R., Hollenbach D.J., Tielens A.G.G.M. 1996, ApJ 466, 561;
\reference Mirabel I.F. 1989, ApJ, 340, L13
\reference Morganti R., Oosterloo T., Tadhunter C.N., van Moorsel, Killeen N.,
  Wills   K.A.  2001, MNRAS 323, 331
\reference Oosterloo T.A., Morganti R., Tzioumis A. et al. 2000, AJ 119, 2085;
\reference Silk J., Rees M.J. 1998, A\&A 331, L1
\reference Spinrad H., Djorgovski S., Marr J., Aguilar L. 1985, PASA 97, 932
\reference Tadhunter C.N., Wills K.A., Morganti R., Oosterloo
T., Dickson R. 2001 MNRAS 327, 227
\reference Tadhunter C.N. et al. 1999, ApJL 512, L91

\end{document}